\documentstyle[titlepage,preprint,aps,floats,epsfig]{revtex}
\newcommand{\beq}{\begin{equation}}
\newcommand{\eeq}{\end{equation}}
\newcommand{\eq}[1]{eq.(\ref{#1})}
\begin{document}
\draft
\preprint{UK/04-11}
\tighten
\title {Three-Loop Reducible Radiative Photon Contributions to  Lamb
Shift and Hyperfine Splitting}
\medskip
\author {Michael I. Eides \thanks{E-mail address:
eides@pa.uky.edu, eides@thd.pnpi.spb.ru}}
\address{Department of Physics and Astronomy,
University of Kentucky,
Lexington, KY 40506, USA\\
and
Petersburg Nuclear Physics Institute,
Gatchina, St.Petersburg 188300, Russia}
\author{Valery A. Shelyuto \thanks{E-mail address:
shelyuto@vniim.ru}}
\address{D. I.  Mendeleev Institute of Metrology,
St.Petersburg 198005, Russia}

\maketitle

\begin{abstract}
Corrections of order $\alpha^3(Z\alpha)^5m$ to the Lamb shift and
corrections of order $\alpha^3(Z\alpha)E_F$ to hyperfine splitting
generated by the insertions of the three-loop one-particle reducible
diagrams with radiative photons in the electron line are calculated.
The calculations are performed in the Yennie gauge.
\end{abstract}


\section{Introduction}

The theory of Lamb shift in light hydrogenlike atoms is rapidly
developing. Calculations of the last corrections which are
significantly larger than 1 kHz for the $1S$ state in hydrogen were
completed recently. These are corrections of orders
$\alpha^3(Z\alpha)^4m$, $\alpha(Z\alpha)^nm$, and
$\alpha^2(Z\alpha)^6m$. The previously unknown correction to the Lamb
shift of order $\alpha^3(Z\alpha)^4m$ is connected with the three-loop
contribution to the slope of the Dirac form factor, and was obtained in
\cite{melrit}.  Corrections of  orders $\alpha(Z\alpha)^nm$ for
$n=4,5,6$ are already well known perturbatively for some time (see,
e.g., review \cite{egs01r}), but relatively large magnitude of the
contributions with $n=6$, and high precision of the experimental data
\cite{nier,debeau} required calculation of the corrections of higher
order in $Z\alpha$. All such corrections were obtained numerically
without expansion in  $Z\alpha$ \cite{jms99,jms01}.  Corrections of
order $\alpha(Z\alpha)^6m$  for higher energy levels were also
calculated recently with high accuracy \cite{jbmis,bjmis}. From the
practical point of view the corrections of order $\alpha(Z\alpha)^nm$
are no more a significant source of theoretical uncertainty for the
Lamb shift, for example for the lowest $S$ and $P$ states in hydrogen
these contributions are now known with uncertainty about $1$ Hz.

Another recent success is connected with the corrections of order
$\alpha^2(Z\alpha)^6m$. The leading logarithm cubed correction of this
order was calculated in the pioneering work \cite{kar93}. This paper
was followed by a heated discussion about the magnitude of the
nonleading logarithmic contributions of order $\alpha^2(Z\alpha)^6m$,
and even the magnitude of the leading logarithm cubed term was put
under suspicion (see, e.g., review \cite{egs01r}). The doubts
about the magnitude of the leading logarithm cubed term were put to
rest in \cite{pach01}, where all logarithmically enhanced terms of
order $\alpha^2(Z\alpha)^6m$ were calculated in the Coulomb gauge.
Finally, the dominant part of the nonlogarithmic contribution of order
$\alpha^2(Z\alpha)^6m$ was obtained in \cite{pachjent03}. At the
present stage remaining uncertainty of the contributions of order
$\alpha^2(Z\alpha)^6m$ to the Lamb shift is about 0.9 kHz and 0.1
kHz for the $1S$ and  $2S$ states in hydrogen, respectively.

The largest still unknown contributions to the Lamb shift in hydrogen
are corrections of order $\alpha^3(Z\alpha)^5m$. The magnitude of these
corrections can be easily estimated multiplying the corrections of
order $\alpha^2(Z\alpha)^5m$ \cite{pach2,esjetp,es} by an extra factor
$\alpha/\pi$. As a result of this simple exercise we see that the
corrections of order $\alpha^3(Z\alpha)^5m$ should be about 1 kHz
for the $1S$ state in hydrogen. In \cite{es03} we calculated radiative
corrections to the Lamb shift of order $\alpha^3(Z\alpha)^5m$  and
radiative corrections to hyperfine splitting of order
$\alpha^3(Z\alpha)E_F$ generated by the diagrams with insertions of
radiative photons and electron polarization loops in the graphs with
two external photons.

\begin{figure}[ht]
\centerline{\epsfig{file=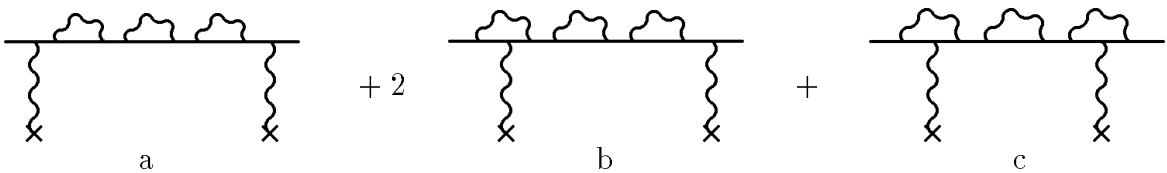,height=2.5cm}}
\vspace{0.5cm}
\caption{Reducible three-loop diagrams}
\label{red}
\end{figure}

Below we calculate in the Yennie gauge corrections of order
$\alpha^3(Z\alpha)^5m$ to the Lamb shift and corrections of order
$\alpha^3(Z\alpha)E_F$ to hyperfine splitting generated by the diagrams
in Fig.\ \ref{red} with insertions of the three-loop one-particle
reducible diagrams with radiative photons in the electron line.

\section{Factorized Contributions to the Lamb Shift and Hyperfine
Splitting}

\subsection{Skeleton Diagram Contributions}

\begin{figure}[ht]
\centerline{\epsfig{file=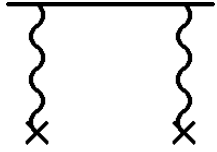,height=2cm}}
\vspace{0.5cm}
\caption{Skeleton two-photon diagram}
\label{skel}
\end{figure}

The diagrams for nonrecoil corrections of order $\alpha^3(Z\alpha)^5m$
to the Lamb shift and corrections of order $\alpha^3(Z\alpha)E_F$ to
hyperfine splitting in Fig.\ \ref{red} can be obtained by
three-loop radiative insertions in the skeleton diagram in Fig.\
\ref{skel}. Respective corrections of lower orders in $\alpha$
generated by one- and two-loop radiative insertions are already well
known (see, e.g., review \cite{egs01r}). All  corrections of order
$\alpha^n(Z\alpha)^5m$ and $\alpha^n(Z\alpha)E_F$ may be calculated in
the scattering approximation (see, e.g., \cite{eksann1}).

Calculation of all these contributions starts with the skeleton diagram
in Fig.\ \ref{skel}. Contribution of each of the diagrams in Fig.\
\ref{red} to the Lamb shift is described by the integral

\beq        \label{nonrecskel}
-\frac{16(Z\alpha)^5}{\pi
n^3}\left(\frac{m_r}{m}\right)^3\:m\int_0^\infty\frac{d|{\bf k}|}
{|{\bf k}|^4}L({\bf k})\:\delta_{l0},
\eeq

\noindent
where $m$ is the electron mass, $M$ is the proton mass, $m_r=m/(1+m/M)$
is the reduced mass, $\alpha$ is the fine structure constant,  $Z$ is
the nucleus charge in terms of the electron charge ($Z=1$ for hydrogen
and muonium), and $|{\bf k}|$ is the magnitude of the dimensionless
momentum of the external photons measured in the units of the electron
mass. The function $L({\bf k})$ describes radiative corrections
to the skeleton diagram, and should be calculated for each particular
diagram in Fig.\ \ref{red}. It is normalized to the contribution of
the skeleton numerator, $L_{skel}({\bf k})=1$. The skeleton
contribution to the Lamb shift is infrared divergent. For some diagrams
in Fig.\ \ref{red} this infrared divergence survives in the Feynman gauge
even after insertion of the factor $L({\bf k})$ which describes
radiative insertions. In order to avoid such spurious infrared
divergencies which anyway cancel in the gauge-invariant sets of
diagrams we use infrared safe Yennie gauge for radiative photons in the
calculations below.

Contribution of each of the diagrams in Fig.\ \ref{red} to hyperfine
splitting is described by the integral\footnote{We
define the Fermi energy $E_F$ as

\beq      \label{muonfermi}
E_{F}=\frac{16}{3}Z^4\alpha^2
\frac{m}{M}(1+a_\mu)\left(\frac{m_r}{m}\right)^{3}ch\:R_{\infty},
\eeq

\noindent
where $m$ is the electron mass, $M$ is the muon mass, $m_r=m/(1+m/M)$
is the reduced mass, $c$ is the velocity of light, $h$ is the Planck
constant, $R_{\infty}$ is the Rydberg constant, and $a_\mu$ is the muon
anomalous magnetic moment.}

\beq        \label{skelhfs}
\frac{8Z\alpha}{\pi n^3}E_F\int_0^\infty \frac{d{|{\bf k}|}}{|{\bf k}|^2}
F({\bf k}),
\eeq

\noindent
where $F({\bf k})$ describes radiative corrections
to the skeleton diagram, and, as the function $L({\bf k})$ in the
case of Lamb shift, should be calculated for each particular diagram in
Fig.\ \ref{red}.  It is normalized to the contribution of the skeleton
numerator, $F_{skel}({\bf k})=1$.

\subsection{Mass Operator and One-Loop Vertex with One
On-Mass-Shell Leg in the Yennie Gauge}

To calculate contributions of the diagrams in Fig.\ \ref{red} we use
explicit expressions for the electron mass operator and the
electron-photon vertex in the Yennie gauge. The one-loop electron
self-energy in the Yennie gauge renormalized on the mass-shell is well
known, and has the form (see, e.g.  \cite{eksann1})

\beq            \label{sigma}
\Sigma(p-k)=-\frac{3\alpha}{4\pi}
(\hat p-\hat k-1)^2(\hat{p}-\hat{k})M({\bf k}),
\eeq

\noindent
where

\beq
M({\bf k})=\frac{1}{1-{\bf k}^2}+\frac{{\bf k}^2}{(1-{\bf k}^2)^2}
\ln {\bf k}^2,
\eeq

\noindent
all momenta are dimensionless, and are measured in the units of
the electron mass $m=1$, and $p_\mu=(1,{\bf 0})$, $pk=0$,
$k^\mu=(0,{\bf k})$.

Renormalized vertex operator with one on-mass-shell leg in the Yennie
gauge has the form \cite{eks93} (we omit terms proportional to the
momentum $k_\mu$ because they do not contribute to the Lamb shift and
hyperfine splitting)

\beq
\label{fyf}
\Lambda_\mu=\frac{\alpha}{2\pi}\{A({\bf k}){\bf k}^2\gamma_\mu
+B({\bf k})\gamma_\mu(\hat p-\hat k-1)
+C({\bf k})p_\mu(\hat p-\hat
k-1)+E({\bf k})\sigma_{\mu\nu}k^\nu\},
\eeq

\noindent
where

\beq
A({\bf k})=-\left(\frac{2}{|{\bf k}|^3}+\frac{1}{2|{\bf
k}|}\right)\Phi({\bf k}) +\frac{2}{{\bf k}^2}S({\bf k})
-\frac{3}{2}M({\bf k})-2\frac{\ln{{\bf k}^2}}{{\bf k}^2}
-\frac{3}{2}\frac{\ln{{\bf k}^2}}{1-{\bf k}^2},
\eeq
\beq
B({\bf k})=-\left(\frac{1}{|{\bf k}|}+\frac{|{\bf
k}|}{8}\right)\Phi({\bf k}) +\frac{1}{2}S({\bf k})-\frac{5}{4}M({\bf
k}) +\frac{1}{4}
-\frac{1}{8}\ln{{\bf
k}^2}-\frac{7}{8}\frac{\ln{{\bf k}^2}}{1-{\bf k}^2},
\eeq
\beq
C({\bf k})=\frac{1}{|{\bf k}|}\Phi({\bf k})-S({\bf k})-\frac{1}{2}M({\bf k})
+\frac{1}{2}\ln{{\bf k}^2}-\frac{3}{2}\frac{\ln{{\bf k}^2}}{1-{\bf k}^2},
\eeq
\beq
E({\bf k})=-\frac{|{\bf k}|}{8}\Phi({\bf k})-\frac{1}{2}S({\bf k})
-\frac{1}{4}M({\bf k})
+\frac{1}{4}+\frac{3}{8}\ln{{\bf k}^2}-\frac{3}{8}\frac{\ln{{\bf k}^2}}
{1-{\bf k}^2},
\eeq

\noindent
and

\beq
\Phi({\bf k})=|{\bf k}|\int_0^1\frac{dz}{1-{\bf k}^2z^2}\ln{\frac{1
+{\bf k}^2z(1-z)}{{\bf k}^2z}}
\eeq
\[
=\rm{Li}(1-|{\bf k}|)-\rm{Li}(1+|{\bf k}|)+2\left[\rm{Li}
\left(1+\frac{\sqrt{{\bf
k}^2+4} +|{\bf k}|}{2}\right)\right.
\left.  -\rm{Li}\left(1-\frac{\sqrt{{\bf
k}^2+4}+|{\bf k}|}{2}\right)-\frac{\pi^2}{4}\right],
\]
\[
\]
\[
S({\bf
k})=\frac{\sqrt{{\bf k}^2+4}}{2|{\bf k}|}\ln{ \frac{\sqrt{{\bf k}^2+4}+|{\bf
k}|}{\sqrt{{\bf k}^2+4}-|{\bf k}|}}.
\]

Euler dilogarithm Li is defined here as in \cite{eksann1}, and the
function $\Phi({\bf k})$ usually arises in calculations of the
diagrams with factorized radiative insertions in the electron line,
see, e.g., \cite{eks1}.

\subsection{Factorized Corrections of Order $\alpha^3(Z\alpha)^5m$ to
the Lamb Shift}

We use explicit expression for the self-energy operator in \eq{sigma},
and calculating the spinor projection on the Lamb shift obtain  the
contribution of the graph $a$ in Fig.\ \ref{red} to the Lamb
shift\footnote{From now on $|{\bf k}|=k$.}

\beq \label{lamb1}
\Delta E_L^{a}= \frac{\alpha^3(Z\alpha)^5}{\pi^2 n^3}
\left(\frac{m_r}{m}\right)^3m
\biggl(-\frac{27}{8\pi^2}\biggr)\int_0^\infty {dk} M^3(k) (1-k^2)
\eeq
\[
= \left(-\frac{135}{1024}\pi^2 + \frac{81}{64}\right)
\frac{\alpha^3(Z\alpha)^5}{\pi^2 n^3}
\left(\frac{m_r}{m}\right)^3m.
\]

Performing similar calculations for the graph $b$ in Fig.\
\ref{red} we obtain

\beq              \label{lamb2}
2\Delta E_L^{b}
=
\frac{\alpha^3(Z\alpha)^5}{\pi^2 n^3}\left(\frac{m_r}{m}\right)^3 m
\biggl(-\frac{9}{2\pi^2}\biggr)\int_0^\infty {dk}
M^2(k)(1-k^2)\,(B(k)+C(k)-E(k))
\eeq
\[
=\left(- \frac{153}{512}\pi^2+\frac{63}{32}\right)
\frac{\alpha^3(Z\alpha)^5}{\pi^2 n^3}
\left(\frac{m_r}{m}\right)^3m.
\]

For the diagram $c$ in Fig.\ \ref{red} we obtain

\beq         \label{lamb3}
\Delta E_L^{c}=
\frac{\alpha^3(Z\alpha)^5}{\pi^2 n^3}\left(\frac{m_r}{m}\right)^3m
\biggl(-\frac{3}{2\pi^2}\biggr) \int_0^\infty {dk} M(k) \biggl\{ k^2
A(k)\Bigl[A(k)-2(B(k)+C(k)-E(k))\Bigr]
\eeq
\[
+ [B(k)+C(k)-E(k)]^2\biggr\}=
-4.~305~82~(1)~
\frac{\alpha^3(Z\alpha)^5}{\pi^2 n^3}
\left(\frac{m_r}{m}\right)^3m.
\]

\subsection{Factorized Corrections of Order $\alpha^3(Z\alpha)E_f$ to
Hyperfine Splitting}

Like in the case of the Lamb shift we use explicit expression
for the self-energy operator in \eq{sigma}, and calculating the spinor
projection on hyperfine splitting obtain  the contribution of the
graph $a$ in Fig.\ \ref{red} to hyperfine splitting

\beq     \label{hfs1}
\Delta E_{HFS}^{a}=\frac{\alpha^3(Z\alpha)}{\pi^2 n^3}\, E_F
\frac{27}{8\pi^2}\int_0^\infty {dk} (1-k^2) \,(2-k^2)\, M^3(k)
\eeq
\[
=\Biggl(\frac{1215}{1024}\pi^2
- \frac{81}{64} ~\Biggr)\frac{\alpha^3(Z\alpha)}{\pi^2 n^3}\, E_F.
\]

Performing similar calculations for the graph $b$ in Fig.\
\ref{red} we obtain

\beq          \label{hfs2}
2\Delta E_{HFS}^{b}
=\frac{\alpha^3(Z\alpha)}{\pi^2 n^3}\, E_F
\frac{9}{2\pi^2}
\int_0^\infty {dk} M^2(k) \,(1-k^2) \Bigl[-\,k^2 A(k)+2B(k)+C(k)-E(k)
\Bigr]
\eeq
\[
=-22.~064~414~(1)~\frac{\alpha^3(Z\alpha)}{\pi^2 n^3}\, E_F.
\]

For the graph $c$ in Fig.\ \ref{red} we obtain

\beq          \label{hfs3}
\Delta E_{HFS}^{c}=
\frac{\alpha^3(Z\alpha)}{\pi^2 n^3}\, E_F \frac{3}{2\pi^2}
\int_0^\infty {dk} M(k)\biggl\{k^2 A(k)\Bigl[k^2 A(k)-2B(k)-C(k)\Bigr]
\eeq
\[
~+~ (2-k^2)\,B(k)\,[B(k)+C(k)-E(k)] + k^2 \,E(k)\,
[B(k)+C(k)-E(k)] ~\biggr\}
\]
\[
=11.~723~748~(6)~
\frac{\alpha^3(Z\alpha)}{\pi^2 n^3}\, E_F.
\]

\section{Summary}

In this paper we calculated factorized corrections
of order $\alpha^3(Z\alpha)^5m$ to the Lamb shift, and factorized
corrections of order $\alpha^3(Z\alpha)E_F$ to hyperfine splitting
generated by the diagrams in Fig.\ \ref{red}. Collecting
contributions to the Lamb shift in \eq{lamb1}, \eq{lamb2},
and \eq{lamb3}, we obtain

\beq
\Delta E^{tot}_{L}=-5.~321~93~(1)~\frac{\alpha^3(Z\alpha)^5}{\pi^2
n^3}\left(\frac{m_r}{m}\right)^3\:m,
\eeq

\noindent
or

\beq \label{totlamb}
\Delta E^{tot}_{L}=-0.~535~\mbox{kHz}
\eeq

\noindent
for the $1S$ level in hydrogen.

Collecting all contributions to hyperfine splitting in
\eq{hfs1}, \eq{hfs2}, and \eq{hfs3} we obtain

\beq     \label{tothfs}
\Delta E^{tot}_{HFS}  =0.~104~23~(1)~
\frac{\alpha^3(Z\alpha)}{\pi^2}\,E_F  ,
\eeq

\noindent
or

\beq
\delta E^{tot}_{HFS} =0.~000~13~~\mbox{kHz}
\eeq

\noindent
for the ground state in muonium.

The result in \eq{totlamb} has just the scale we expected on the basis
of the general considerations  explained in the Introduction, and
corrections of this  magnitude are phenomenologically relevant at the
current level of the experimental and theoretical accuracy (see, e.g.
\cite{egs01r}). Work on calculation of nonfactorizable contributions is
now in progress, and we postpone discussion of the phenomenological
implications of the results in \eq{totlamb} and \eq{tothfs} until its
completion.

\acknowledgements

This work was supported by the NSF grant PHY-0138210. The work
of V.  A. Shelyuto was also supported in part by the RFBR grant
03-02-16843 and DFG grant GZ 436 RUS 113/769/0-1.

\end{document}